# On implementation of ferrite magnetostatic/magnetoelectric particles for quantum computation

E.O.Kamenetskii [1] and O. Voskoboynikov [2]

[1] Department of Electrical and Computer Engineering,
Ben Gurion University of the Negev, Beer Sheva, 84105, Israel

[2] Department of Electronics Engineering & Institute of Electronics,
National Chiao Tung University, Hsinchu, Taiwan 3000, R.O.C.

We consider an implementation of quantum gates for quantum computation using magnetostatic/magnetoelectric (MS/ME) macroscopically quantized states in small ferrite disks. Confinement phenomena for MS oscillations in a normally magnetized ferrite disk show typical atomic properties like discrete energy levels. Because of discrete energy eigenstates of MS oscillations, the oscillating system is described as a collective motion of quasi-particles – the light magnons. A macroscopic quantum analysis of MS oscillations underlies the physics of quantized ME oscillating spectrums in ferrite disks with surface electrodes. We discuss possible technologies for physical realization of new logic gates based on MS/ME-particle qubits.

## 1. INTRODUCTION

The field of quantum computation has been stimulated by fundamental discovery that devices using unique quantum-mechanical features can perform information processing. Various physical implementations are being explored at present for quantum computation. Among them there are, in particular, solid-state systems based on semiconductor quantum dots [1,2]. In this paper we discuss the possibility to use quantized states in another-type solid-state systems – small ferrite-disk particles with the so-called magnetostatic/magnetoelectric (MS/ME) oscillating spectra – for quantum computation.

Can ferrite particles be considered as "magnetic quantum dots" – a certain analog of semiconductor quantum dots? In a series of new publications, confinement phenomena of high-frequency magnetization dynamics in magnetic particles have been the subject of much experimental and theoretical attention (see e.g. [3]). Mainly, these works are devoted to the important studies of the magnetization spectra, but do not focus on the energy eigenstates of a whole ferrite–particle system. Till now there were no (to the best of our knowledge) physical models of small ferrite particles with high-frequency magnetization dynamics, which demonstrate and analyze the *energy eigenstates of a whole ferrite-particle system* similarly to semiconductor quantum dots.

Magnetostatic (MS) oscillation in ferrite resonators have the wavelength much smaller than the electromagnetic wavelength at the same frequency and, at the same time, much larger than the exchange-interaction spin wavelength. This intermediate position between the "pure" electromagnetic and spin-wave processes reveals very special behaviors of the geometrical effects. Necessary condition for excitation of multiple MS modes in a ferrite *ellipsoid* (or a ferrite *sphere*, as a particular case) is that the exciting RF magnetic field at the sample be essentially *non-uniform*. In this case, however, one have just only a very few (and rather broad) absorption peaks in a spectrum [4]. The situation can be completely different when *non-ellipsoidal* (*non-spherical*) ferrite samples are used. In such a case, one can see a long series of oscillating MS modes excited by the *uniform* RF magnetic field. For a normally magnetized planar straight-edge ferrite resonator one observes multi-resonance, but strongly *irregular*, absorption spectrums [5]. Striking difference in the picture of absorption spectrums one can discern in a case of a normally magnetized planar *disk* ferrite resonator. There are the *multi-resonance regular* experimental spectrums with properties similar to an *atomic-like* $\delta$ - function density of states [6,7]. The character of the multi-resonance spectra (obtained with respect to a DC bias magnetic field) leads to a clear conclusion that the energy of a source of a DC magnetic field is absorbing "by portions" or discretely, in other words. What can be the nature of such a strong discreteness of the ferrite-disk-resonator spectrum? Certainly, there should be some inner mechanism of *quantization* of the DC energy absorbed by a small disk-form ferrite sample. As it was shown in [8], there is an essential difference between the orthonormality relations for MS modes in a ferrite sphere and in a normally magnetized ferrite disk. Just for a normally magnetized flat ferrite disk we can obtain a complete *discrete spectrum of energy levels*. In this case the MS potential function can be considered as a *probability distribution function*. For such a complex scalar function, the wave equation is the Schrödinger-like equation. This fact allows analyzing the MS oscillations as macroscopically quantum mechanical problems and gives a basis for a clearer understanding the nature of the observed multi-resonance spectrum.

A macroscopic quantum analysis for MS oscillations underlies the physics of *magnetoelectric* (ME) oscillating spectrums. Recently, an idea that MS-wave ferrite resonators with special-form surface metallizations can exhibit in microwaves properties of a local particle with internal ME coupling was put forth [9]. Experimental investigations have verified the local microwave ME effect, which is characterized by a multi-resonance spectrum of quasistatic oscillations excited by the external uniform RF electric and magnetic fields and their combination [7]. The spectral pictures demonstrate the unified process of ME oscillations, which have special symmetry properties. Different-symmetry (left-hand or right-hand) ME particles are distinguished by the energy levels.

In this paper we consider a quantum-gate mechanism based on macroscopically quantized MS/ME oscillations. We discuss possible technologies for physical realizations of new logic gates.

## 2. ENERGY EIGENSTATES OF MAGNETOSTATIC OSCILLATIONS IN A NORMALLY MAGNETIZED FERRITE DISK

Confinement phenomena for MS oscillations in a normally magnetized ferrite disk show typical atomic properties like discrete energy levels. The main feature of multi-resonance atomic-like spectra shown in [6,7] is the fact that high-order peaks correspond to lower quantities of the bias DC magnetic field. Physically, the situation looks as follows. Let $H_0^{(A)}$ and $H_0^{(B)}$ be, respectively, the upper and lower values of a bias magnetic field corresponding to the borders of a region. We can estimate a total depth of a "potential well" as: $\Delta U = 4\pi M_0 \left( H_0^{(A)} - H_0^{(B)} \right)$, where $M_0$ is the saturation magnetization. Let $H_0^{(1)}$ be a bias magnetic field, corresponding to the *main absorption peak* in the experimental spectrum ($H_0^{(B)} < H_0^{(1)} < H_0^{(A)}$). When we put a ferrite sample into this field, we supply it with the energy: $4\pi M_0 H_0^{(1)}$. To some extent, this is a pumping-up energy. *Starting from this level,* we can excite the entire spectrum from the main mode to the high-order modes. As a value of a bias magnetic field decreases, the "particle" obtains the *higher levels of negative energy*. One can estimate the negative energies necessary for transitions from the main level to upper levels. For example, to have a transition from the first level $H_0^{(1)}$ to the second level $H_0^{(2)}$ ($H_0^{(B)} < H_0^{(2)} < H_0^{(1)} < H_0^{(A)}$) we need the density energy surplus:

$$\Delta U_{12} = 4\pi M_0 \left( H_0^{(1)} - H_0^{(2)} \right). \tag{1}$$

The situation is very resembling the increasing a negative energy of the hole in semiconductors when it "moves" from the top of a valence band. In a classical theory, negative-energy solutions are rejected because they cannot be reached by a continuous loss of energy. But in quantum theory, a system can jump from one energy level to a discretely lower one; so the negative-energy solutions cannot be rejected, out of hand. When one continuously varies the quantity of the DC field $H_0$, for a given quantity of frequency $\omega$, one sees a discrete set of absorption peaks. It means that one has the discrete-set levels of potential energy. The line spectra appear due to the quantum-like transitions between energy levels of a ferrite disk-form particle. As a quantitative characteristic of permitted quantum transitions, there is the probability, which define the intensities of spectral lines. The quantized-like transitions for MS oscillations in a normally magnetized thin-film ferrite disk were demonstrated in recent experimental studies [10].

A magnetically ordered sample is, in principle, a quantum mechanical system. So one cannot be surprised that for certain collective-mode processes (in our case, the dipole-dipole-interaction MS-wave processes) and a certain geometry (in our case, a normally magnetized ferrite disk) the *macroscopic quantum bound states* can be observed. MS oscillations are described by *scalar wave function* $\psi$. In a case of a lossless axially magnetized ferrite rod this scalar wave function should satisfy the following equation [11]:

$$a^{(1)}(z)\frac{\partial^2 \psi(z,t)}{\partial z^2} + a^{(2)}(z)\psi(z,t) = \frac{\partial \psi(z,t)}{\partial t} \tag{2}$$

This is the stationary-state Schrödinger-like equation. For a MS-wave waveguide based on an axially magnetized ferrite cylinder one has a *complete-set discrete spectrum of propagating MS modes* [8,12]. For a monochromatic process ($\psi \sim e^{i(\omega t - \beta z)}$), the $\psi$ function is expanded by the complete-set membrane functions $\widetilde{\varphi}$ of MS-wave waveguide modes. In this case we have an infinite set of differential equations [everyone is similar to Eqn. (2)] written for waveguide modes. For frequency $\omega$ and for a certain $n$-th waveguide mode, we obtain from Eqn. (2):

$$-a_n^{(1)}\beta_n^2 + a_n^{(2)} = i\omega \tag{3}$$

For harmonic processes, coefficients $a^{(1)}$ and $a^{(2)}$ should be imaginary quantities.

In a normally magnetized ferrite-disk resonator with a small thickness to diameter ratio, the monochromatic MS-wave potential function $\psi$ is represented as:

$$\psi = \sum_{p,q} A_{pq} \widetilde{\xi}_{pq}(z) \widetilde{\varphi}_q(\rho,\alpha), \tag{4}$$

where $A_{pq}$ is a MS mode amplitude, $\tilde{\xi}_{pq}(z)$ and $\tilde{\varphi}_q(\rho,\alpha)$ are dimensionless functions describing, respectively, "thickness" ($z$ coordinate) and "in-plane", or "flat" (radial $\rho$ and azimuth $\alpha$ coordinates) MS modes. For MS "flat" functions $\tilde{\varphi}$ in a ferrite-disk resonator one can formulate *the energy eigenvalue problem* [8]:

$$\hat{F}_\perp \tilde{\varphi}_q = E_q \tilde{\varphi}_q, \tag{5}$$

where $\hat{F}_\perp$ is a two-dimensional ("in-plane") differential operator. The normalized energy of MS mode $q$ in a thin-film ferrite disk is expressed as:

$$E_q = \frac{1}{2} g \mu_0 (\beta_q)^2, \tag{6}$$

where $g$ is the unit dimensional coefficient, $\beta_q$ is the mode wavenumber. The energy orthonormality in a ferrite disk is described as:

$$(E_q - E_{q'}) \int_Q \tilde{\varphi}_q \tilde{\varphi}_{q'}^* dQ = 0, \tag{7}$$

where $Q$ is a cylindrical cross section of an open disk. In our description of MS oscillations we neglect the exchange interaction and the "magnetic stiffness" is characterized by the "weak" dipole-dipole interaction. Because of discrete energy eigenstates of MS-wave oscillations resulting from structural confinement in a special case of a normally magnetized ferrite disk, one can describe the oscillating system as a collective motion of quasi-particles – the "light magnons" (lm). The meaning of this term arises from the fact that effective masses of these quasi-particles is much less than effective masses of "heavy" magnons – the quasi-particles existing due to the exchange interaction. For a monochromatic MS-wave mode, the light-magnon effective mass is expressed as [11]:

$$\left(m_{eff}^{(lm)}\right)_q = \frac{\hbar}{2} \frac{\beta_q^2}{\omega} \tag{8}$$

This expression looks very similar to an effective mass of the "heavy" magnon for spin waves with the quadratic character of dispersion [13]. The light magnons have the *reflexively-translational motion* between the lower $(z=0)$ and upper $(z=b)$ planes of a ferrite disk.

## 3. MAGNETOELECTRIC OSCILLATIONS

A macroscopic quantum analysis for MS oscillations [8,11] underlies the physics of *magnetoelectric* (ME) oscillating spectrums observed in ferrite disks with surface electrodes [7]. Recent theoretical analysis and new experimental results show that due to special motion processes of light magnons, MS oscillations in a normally magnetized ferrite disk are characterized by the eigen electric moments [14]. This gives a clue to the experimentally observed ME effect in small ferrite resonators with special-form surface metallizations.

One of the main feature of experimental spectra [7], obtained for a linear surface electrode placed on a disk resonator (see Fig. 1 a), is the fact that for different-type RF fields of the cavity (electric, magnetic, and combined) we see the same positions of the absorption peaks with respect to bias magnetic field $\vec{H}_0$. It means that different types of fields (the $\vec{E}$ - or the $\vec{H}$ - fields) excite the same spectrum of ME oscillation modes. However, effectiveness of excitation of different ME modes in the spectrum (mode amplitudes) depends on a type of the exciting fields. Since different types of the exciting fields produce the same oscillation spectrum, a system is characterized by a *set of parameters* with certain spectral properties. This fact gives us a possibility to represent a disk-type ME particle as a particle characterized by two (electric $\vec{p}_e$ and magnetic $\vec{p}_m$) dipole moments. The moments are related to external RF $\vec{E}$ and $\vec{H}$ fields as:

$$\vec{p}_e = \vec{\vec{\alpha}}_{ee} \vec{E} + \vec{\vec{\alpha}}_{em} \vec{H}$$
$$\vec{p}_m = \vec{\vec{\alpha}}_{me} \vec{E} + \vec{\vec{\alpha}}_{mm} \vec{H} , \qquad (9)$$

where dyadic polarizabilities $\vec{\vec{\alpha}}_{ee}, \vec{\vec{\alpha}}_{em}, \vec{\vec{\alpha}}_{me}$, and $\vec{\vec{\alpha}}_{mm}$ are defined from ferrite material properties and geometry of a disk-type ME particle. These polarizabilities are *parameters* of a system, which are characterized by certain spectral properties (certain regular positions of poles and zeros with respect to frequency or bias magnetic field). Because of these spectral properties one can talk about *the unified process of ME oscillations*. For different-type RF fields in the cavity we have certain stationary states of ME oscillations.

In papers [7] it was showed that a disk-type ME particle is characterized as a "glued pair" of two (electric $\vec{p}_e$ and magnetic $\vec{p}_m$) mutually perpendicular dipole moments. The observed ME oscillations have special symmetry properties. This fact can be illustrated by a simple model. The particle is considered as a triple of three mutually perpendicular vectors: an electric dipole with a moment $\vec{p}_e$ (a polar vector), a magnetic dipole with a moment $\vec{p}_m$ (an axial vector) and bias magnetic field $\vec{H}_0$ (an axial vector). One may have the left-hand or the right-hand particles, which are characterized, respectively, by the left-hand or the right-hand triples of vectors $\vec{p}_e, \vec{p}_m$ and $\vec{H}_0$ (Fig. 1 a). As we discussed in [15], the PT invariance (the time-reversal operation T combined with the parity P) does not hold in this model (Fig. 1 b), but the CPT invariance (when the charge conjugation C changes a sign of a vector $\vec{p}_e$) takes place (Fig. 1 c) [16]. The problem of the charge conjugation has to be considered in connection with a sign of energy eigenstates of ME oscillations in a particle. Different-symmetry (left-hand or right-hand) ME particles are distinguished by the energy levels. In an absorption spectrum of ferrite ME obtained in the external RF electric field [7], the two adjacent peaks (say the first and the second peaks) should be characterized by different symmetry properties.

## 4. MS/ME PARTICLES AS QUBITS

Our efforts here focus on the possible implementation of the observed macroscopically quantized MS/ME oscillations for quantum computation. To achieve the conditions for quantum computation, it is required to have precession control of Hamiltonian operations on well-defined two-level quantum systems and a very high degree of quantum coherence. It is also required to have a good entanglement between different parts of the quantum computer and, at the same time, a bad entanglement between the quantum computer and its environment [17]. A small ferrite disk with MS/ME oscillations can be considered as a complex system whose evolution is both coherent (i.e. unitary) and controllable. The whole ferrite particle is the Hamiltonian system. Because of the possibility to formulate the energy eigenvalue problem for MS modes in a normally magnetized ferrite disk, one has the complete-set energy eigenstates and the Hilbert functional space of orthonormal eigenfunctions. Decoherence time is defined from the breadth of a spectral line. In our case this breadth is due to intrinsic relaxation processes in a ferrite material and the radiation damping of a whole ferrite sample. Based on a classical electromagnetic theory of a precessing point magnetic dipole, it was shown in [18] that magnetic-dipolar radiation damping can be a primary source of the broadening of the *uniform mode* in ferromagnetic resonance. This classical mechanism cannot give, however, any explanations for possible line broadening for non-uniform modes, which we observe in the multi-resonance spectrum. The radiation damping for these modes should be considered as a problem involving discrete bound states and may take place only due to the quantized picture. External interactions increasing the probability of transition of a quantum system to other states lead to line broadening. With use of a special cavity construction we are able to enhance this interaction [10]. On the other hand, for experiments with a high-$Q$ cavity [7] the ferrite-particle-cavity-field entanglement can be strongly reduced. The minimum line breadth we can estimate as the width due to intrinsic relaxation processes in a magnetically ordered material. The minimum width of a ferromagnetic resonance peak for YIG at a room temperature is about $\Delta H_0 = 0.3\, Oe$ [13]. Based on this value one easily estimates that the quantities of the decoherence time in MS/ME particles, $\tau_\phi$, can be on the order of microseconds. These quantities of the decoherence times in a ferrite particle *at a room temperature* are on the same order as the spin phase decoherence times in GaAs quantum dots at *law temperatures* [2]. This spin decay is generally a source of decoherence, i.e. loss of quantum information encoded in the qubit. The minimum time required to execute one quantum gate, $\tau_{switch}$, is estimated as $\hbar/\Delta E$, where $\Delta E$ is the energy splitting in the two-level system. The ratio $\tau_\phi/\tau_{switch}$ gives the largest number of steps permitted in a computation using qubits. The estimations made in [11] show that in our case (because of a big volume of a particle, where the macroscopically quantized MS/ME oscillations take place) the energy splitting $\Delta E$ is extremely big and, therefore, the ratio $\tau_\phi/\tau_{switch}$ becomes extremely big compared to other known qubit systems.

Variation of DC magnetic field $H_0$ leads to variation of permeability tensor $\ddot{\mu}$. So for a discrete set of the $H_0$ quantities, corresponding to sharp field-dependent resonances, we have discrete states of the permeability tensor properties. In a ferromagnetic resonance one can vary parameters of tensor $\ddot{\mu}$ by two ways: (a) by variation of $H_0$ (at constant $\omega$) or (b) by variation of $\omega$ (at constant $H_0$) [13]. The above energy eigenvalue problem was formulated for a case of constant $\omega$. It is not difficult to show that in a case of variation of $\omega$ (at constant $H_0$), the eigenfunctions constitute the orthonormal functional basis and, as a result, one has the same discrete spectrum of energies $E_q$ as well.

As we discussed above, the $\psi$ function can be expanded by complete-set membrane functions $\tilde{\varphi}$ of MS-wave waveguide modes for a monochromatic process ($\psi \sim e^{i\omega t}$). In this case we have an infinite set of differential equations [everyone is similar to Eqn. (2)] written for waveguide modes. This is not the situation one may see in a ferrite disk resonator. Let us consider the case of a constant-value bias magnetic field $H_0$ and varying frequency $\omega$. Every resonance mode in a ferrite disk resonator is described by Eqn. (2). However, since every resonance peak is characterized by its own frequency, one can suppose that there are no complete-set membrane functions $\tilde{\varphi}$ of MS modes in this case. For a given quantity of bias magnetic field $H_0$, let us introduce the following density matrix. $\Theta(\omega_a, \omega_b)$ written for the "in-plane" functions $\tilde{\varphi}$:

$$\Theta(\omega_a, \omega_b) = \int_Q \widetilde{\varphi}^*(\omega_a) \widetilde{\varphi}(\omega_b) dQ, \tag{10}$$

where $\omega_{a,b}$ are frequencies of some two resonance peaks (the peaks numbered as *a* and *b*). Evidently, density matrix $\Theta(\omega_a, \omega_b)$ is characterized by the Hermitian property:

$$\Theta^*(\omega_a, \omega_b) = \Theta(\omega_a, \omega_b) \tag{11}$$

Diagonal elements of the density matrix are defined as:

$$\Theta(\omega_a, \omega_a) = \int_S |\widetilde{\varphi}(\omega_a)|^2 ds \tag{12}$$

Since every resonance "in-plane" function is characterized by its number we can rewrite Eqn. (10) as [19]:

$$\Theta_{m,n}(\omega_m, \omega_n) = \int_S \widetilde{\varphi}_m^*(\omega_m) \widetilde{\varphi}_n(\omega_n) ds \tag{13}$$

A priori, there is no foundation to state that for a given quantity of bias magnetic field $H_0$, when frequencies $\omega_m$ and $\omega_n$ are different, functions $\widetilde{\varphi}_m$ and $\widetilde{\varphi}_n$ are mutually orthogonal. There is, however, a complete correspondence in the resonance peak positions for two types of spectrums: obtained by variation of $H_0$ (at constant $\omega$) or by variation of $\omega$ (at constant $H_0$) [19]. This matching is illustrated in Figs. 2 for first three peaks ($q$=1,2,3). In fact, we are able to place sequentially every peak from the frequency spectrum at the same frequency $f'$ by a sequence of the DC magnetic field values.

For the situation shown in Fig. 2 (when we sequentially place every peak from the frequency spectrum *at the same frequency* ($\omega_m = \omega_n = \omega'$) by a sequence of the DC magnetic field values), it is evident that functions $\widetilde{\varphi}_m$ and $\widetilde{\varphi}_n$ are mutually orthogonal. Really, in accordance with the above consideration, any "in-plane" eigenfunctions are mutually orthogonal for the monochromatic process. Because of a complete correspondence in the resonance peak positions for two types of spectrums, in a case of variation of $\omega$ (at constant $H_0$), the eigenfunctions constitute the orthonormal functional basis and, as a result, one has the same discrete spectrum of energies $E_q$ as well.

We have shown that the discrete transitions between quantum levels in a ferrite particle are possible by two ways: (a) by variation of DC magnetic field at a constant cavity resonance frequency or (b) by variation of a cavity resonance frequency at a constant DC magnetic field. The entire spectrum of MS oscillations in a ferrite disk (from the main peak and up to the high-order peaks) can be excited in a cavity or by a sequence of a decreasing DC magnetic field values at a constant cavity frequency, or by a sequence of increasing cavity frequencies at a constant DC magnetic field. In both cases we have the same *discrete set of parameters* of tensor $\ddot{\mu}$. The "flat-function" alternative magnetization, $\widetilde{m}$, in a lossless normally magnetized disk is expressed by the Landau-Lifshitz equation:

$$\frac{\partial \tilde{\vec{m}}}{\partial t} + \gamma \tilde{\vec{m}} \times \vec{H}_0 = \gamma \vec{M}_0 \times \nabla_\perp \tilde{\varphi} \tag{14}$$

We can see that this phenomenological equation, resulting from the quantum-mechanical exchange stiffness, has stationary states (or with respect to a DC magnetic field, of with respect to frequency). Such stationary states are caused, however, not by the heavy-magnon dimensional quantization (since the sizes of a disk are much less than the exchange-interaction scales), but by the light-magnon dimensional quantization in a ferrite disk. We have a case of the so-called *conditional quantum dynamics* [1]. In our situation the heavy-magnon quantum subsystem undergoes a coherent collective-mode evolution that depends on (and is determined by) the quantum states of another subsystem – the light-magnon subsystem in a ferrite sample.

On the other hand, interaction of RF cavity fields with disk MS/ME particles, shown in experiments [6,7], should also mean a quantum nature. Bound "magnetic" and electric charges in "in-plane" magnetic and electric dipoles are characterized by certain quantized coupling energies. Alternative cavity fields create these quantized bound-charge systems. So the resonance-peak absorption can be described by an annihilation operator for the cavity field. If a MS/ME particle, placed in a cavity, is found initially in an excited state, one can consider a creation operator for the cavity field.

When a MS/ME particle is placed in tangential homogeneous RF magnetic field $\vec{H}_{RF}$, the Hamiltonian is expressed as

$$\hat{H} = \hat{F} + k_1 \hat{\vec{p}}_m \cdot \vec{H}_{RF}, \tag{15}$$

where $\hat{\vec{p}}_m$ is operator corresponding to the "in-plane" magnetic-dipole polarization $\vec{p}_m = \int_Q \tilde{\vec{m}} \, dQ$, $k_1$ is a coefficient.. In a case of a ME particle placed in a tangential RF electric field $\vec{E}_{RF}$, the Hamiltonian is expressed as

$$\hat{H} = \hat{F} + k_2 \hat{\vec{p}}_e \cdot \vec{E}_{RF}, \tag{16}$$

where $\hat{\vec{p}}_e$ is operator corresponding to the electric-dipole polarization, $k_2$ is a coefficient. The energy difference between the first and the second peaks $\Delta E_{12}$, found based on Eqn. (6), is much less than the splitting $\Delta U_{12}$ [11]. From Eqn. (15) one can approximately estimate for RF magnetic field:

$$\Delta U_{ij} V = k_1 \overline{(\vec{p}_{m_i} - \vec{p}_{m_j}) \cdot \vec{H}_{RF}}, \tag{17}$$

where $V$ is a sample volume.

We should consider a MS particle as a semiclassical-semiquantum object. It can be suppose that the *main resonance* (the first absorption peak) in a MS particle appears due to pure classical mechanisms of magnetic polarization. There is, for example, the case of homogeneous precession in a ferrite disk, which is excited by a homogeneous RF magnetic field. The magnetic-dipole moment of a ferrite disk corresponding to the first resonance, $\vec{p}_{m_1}$, can be calculated based on classical models [13]. Further sharp resonance peaks appear due to a macroscopically quantum nature. With variation of $H_0$ (at $\omega$ constant) or variation of $\omega$ (at $H_0$ constant) we have quantized jumps to high-order states. In a ME particle, all the electric-polarization peaks, including the first one, have a macroscopically quantum nature. The electric-resonance peaks are strongly correlated with the magnetic-resonance peaks. On the spectral pictures we have complete coincidence of the electric and magnetic peak positions. Based on experimental data [7] we can conclude that for a disk ME particle with a linear electrode, the following relation takes place:

$$\Delta U_{12} V = 4\pi M_0 \left(H_0^{(1)} - H_0^{(2)}\right) V = k_1 \overline{(\vec{p}_{m_1} - \vec{p}_{m_2}) \cdot (\vec{H}_{RF})_{\max}} = k_2 \overline{(\vec{p}_{e_1} - \vec{p}_{e_2}) \cdot (\vec{E}_{RF})_{\max}} = k_1 \overline{(\vec{p}_{m_1} - \vec{p}_{m_2}) \cdot \vec{H}_{RF}} + k_2 \overline{(\vec{p}_{e_1} - \vec{p}_{e_2}) \cdot \vec{E}_{RF}} \quad (18)$$

where $(\vec{H}_{RF})_{\max}$ and $(\vec{E}_{RF})_{\max}$ are, respectively, maximum magnetic and electric RF fields in a cavity. Similar relations we can write for transition from the main state to any *j*th state.

## 5. REALIZATIONS OF QUANTUM GATES BASED ON MS/ME PARTICLES

The principle of unitary time evolution is the cornerstone of the quantum theory. This principle forms the basic idea for the quantum computer construction. The basic elements of the computer (i.e. qubits) are the MS/ME particles themselves. Because of the quantized states of magnetization we have the quantized states of magnetic dipoles (for a whole ferrite disk) controlled by the quantized states of the light-magnon oscillations. On the other hand, in a ferrite ME particle the quantized states of the light-magnon oscillations control also the quantized states of electric dipoles. Based on these properties of MS/ME particles, we outline the following possible experimental realizations of the quantum controlled-NOT gate.

**(A) ME particle placed in a high-Q cavity**

The first technology is that of a ME particle placed in a center of a high-*Q* microwave rectangular cavity with orientation shown in Fig. 3. We have a bias magnetic field and the disk axis directed along *x*-axis and a linear surface electrode directed along *y*-axis. Cavity quantum electrodynamics (both in the microwave and optical regimes) demonstrates a strong interaction of (natural) atom beams with a single quantized field mode of a cavity [20,21]. In our case there is also the case of "atom"-cavity-mode coupling. We have a system which consist of a discrete-mode cavity field interacting with a quantum system – the ME particle. The target qubit is a cavity, the control qubit is the quantized ME field. Our scheme is based on the adiabatic transfer of ME-particle-state coherence to the cavity mode. In other words, coherence of the ME-particle energy levels is mapped directly onto a cavity field. The initial eigenstates describe the ME-particle-cavity system. The final eigenstates contains a contribution from the excited ME-particle state.

Let $f_1$ be a resonance frequency of cavity mode $TE_{101}$ and $f_2$ be a resonance frequency of cavity mode $TE_{201}$. It is clear that a ME particle placed in a cavity center may interact with the cavity electric field in a case of $TE_{101}$ mode and with the cavity magnetic field in a case of $TE_{201}$ mode. The high-*Q* microwave cavity with these two resonance frequencies can be viewed, respectively, as a two-state system: $|0\rangle$ and $|1\rangle$. The particle characteristic dimension is much less than $\lambda_2 = c/f_2$ (in Fig. 3 we premeditatedly increased the particle sizes for clearer observation). The main absorption peak (the first-order ME mode) corresponds to the "ground" $|g\rangle$ state of a ME particle and the second absorption peak (the second-order ME mode) is the "excited" $|e\rangle$ state. Suppose that we realized a cavity with frequencies $f_1$ and $f_2$ corresponding to frequencies of the ME-particle first two peaks at DC magnetic $H_0^{(I)}$. The initial state $|\psi_{initial}\rangle = |g,0\rangle$ describing the ME-particle-cavity system corresponds to the case of the main ME mode at frequency $f_1$, bias magnetic field $H_0^{(I)}$ and cavity mode $TE_{101}$ (Fig. 4 (a)). For a certain magnetic field $H_0^{(II)} < H_0^{(I)}$ we can shift the oscillating spectrum of the particle to the position shown in Fig. 4 (b). In this case a RF electric field of cavity mode $TE_{101}$ excites the second-mode ME oscillation in a particle. This is the $|e,0\rangle$ state. Now we turn a DC magnetic field back to quantity $H_0^{(I)}$ (Fig. 4 (c)). The final cavity eigenstate ($TE_{201}$ mode) contains a quantized contribution from the excited ME-particle state. As a result, we have a "shift" through the transformations $|g,0\rangle \rightarrow |e,0\rangle \rightarrow |g,1\rangle$. The above mechanism shows how the source (the control qubit – the ME particle) can "teach" the field to evolve toward a desirable quantum state. To a certain extent, we have a situation resembling the cases in [20,21], where it is shown how coherence of Zeeman levels can be mapped onto a cavity field directly. Intensity of the created cavity mode can be increased when instead of one ME particle shown in Fig. 3 we use a set of properly oriented coupled ME particles placed in a center of a cavity. The problem of coupled MS/ME particles we discuss below.

In the above technology we have the quantum entanglement between the external (cavity mode) and the internal (ME normal mode) state of the ME particle placed in high-$Q$ microwave cavity. It gives the possibility to prepare an arbitrary entangled state of the form

$$|\psi_{target}\rangle = \alpha |\psi^e, e\rangle + \beta |\psi^g, g\rangle, \qquad (19)$$

where

$$|\psi^e\rangle = \sum_{n=0}^{1} w_n^e |n\rangle$$

$$|\psi^g\rangle = \sum_{n=0}^{1} w_n^g |n\rangle \qquad (20)$$

are two normalized cavity states associated with the excited and ground states respectively; $\alpha$ and $\beta$ are weighting parameters that control the relative probability and phase between the ground and excited states; $w_n^e$ and $w_n^g$ are the probability amplitudes. The strategy for creating target state (19) is based on the unitary time evolution generated by Hamiltonians (15) and (16).

**(B) Magnetically coupled ME particles**

Our second proposal for the implementation of the quantum controlled-NOT relies on the magnetic dipole-dipole interactions between two qubits. For the purpose of this model the qubits are magnetic dipoles of MS/ME particles. The principal properties of magnetically coupled MS/ME particles follow from the model of two disks side by side in a plane or on top of each other (respectively, lateral or vertical MS/ME molecules).

Let us focus our attention to the case of two laterally coupled MS/ME particles. In our analysis we assume that the two particles are coupled in a quantum-mechanically coherent way. This means that the light-magnon states extended over the whole quantum-particle molecule whereas states localized to the left or right particle do not exist in this description. In our model, when interparticle distance $d$ is equal to zero the "molecule" degenerates to the simple disk MS/ME particle. The energy spectrum of such a disk particle is known [8,11]. For large interparticle distances $d$ we obtain the same spectrum, but twofold now, since the system consists of two identical but completely separated quantum particles. By lowering $d$ starting from the weak-coupling limit we have the properties of a diatomic molecule: the energies decrease or increase with respect to the reference level due to formation of even-parity or odd-parity states [22].

Two laterally coupled ferrite disk particles are depicted in Fig. 5 a. Similar to our previous model of an open ferrite disk [8] we neglect the magnetostatic fields in corner regions. We also neglect non-homogeneity of the internal DC magnetic field. The model is relevant for disks with small thickness-to-diameter ratios. In this case one can use separation of variables and obtain analytical solutions for "flat" MS functions. Because of separation of variables, one can impose independently the boundary conditions – the continuity conditions for the MS potential $\psi$ and for the normal components of the magnetic flux density – on lateral cylindrical surfaces and plane surfaces ($z=0$, $z=b$). As a result, we have to solve a system of two equations: (a) characteristic equation for MS waves in a normally magnetized ferrite slabs and (b) characteristic equation for MS waves in two parallel axially magnetized ferrite cylinders.

MS-waves in two parallel axially magnetized ferrite cylinders are described by Walker's equation:

$$\mu \left( \frac{\partial^2 \psi}{\partial x^2} + \frac{\partial^2 \psi}{\partial y^2} \right) + \frac{\partial^2 \psi}{\partial z^2} = 0 , \qquad (21)$$

where $\mu$ is a diagonal component of the permeability tensor. An analysis should be based on the use of bipolar coordinates. In such a coordinate system one obtains an analytical solution for the Laplace equation [23]. Since for "in-plane" MS functions the Worker equation (21) is, in fact, the modified Laplace equation, one can obtain an analytical solution in our case as well.

MS oscillations in a separated ferrite disk resonator are considered as certain quasiparticles – the light magnons – having quantization of energy and characterizing by certain effective masses. The states of the light magnons are described based on the so-called translational eigenfunctions. The types of light magnons are classified with respect to the azimuth distribution [11]. For a case of a "diatomic molecule" shown in Fig. 5 a, the even-parity and odd-parity energy states are characterized by translational motion (between planes $z=0$ and $z=b$) of even-parity and odd-parity light magnons. The states of a coupled-disk system are entangled: one cannot decompose the even-parity and odd-parity light magnons as light magnons of separated disks.

For laterally coupled normally magnetized ME particles one can easily realize the mechanism of switching on or switching off the coupling. Under one of the particles we place an in-plane small wire loop. The particles have identical parameters and are put in the same homogeneous bias magnetic field. At the absence of a current in a wire loop we have the double-particle system. When a current in a loop is thrown, a bias magnetic field of the particle is changed and magnetic parameters of this particle are changed as well. So the two particles become strongly non-identical. In this case one obtains a vanishing overlap and, therefore, switching off the coupling.

As our emphasis is on ME effects in coupled ferrite particles, we should consider also the property of electric-dipole orientations when every ferrite disk has a linear surface electrode. The mutual orientation of a triple of vectors $\vec{p}_e, \vec{p}_m$ and $\vec{H}_0$ for an even-parity state is shown in Fig. 5 b and for an odd-parity state – in Fig. 5 c. It is clear that by lowering $d$ (starting from the weak-coupling limit) and formation of the properties of a diatomic molecule, we should not have a transformation from the left-hand triple of vectors $\vec{p}_e, \vec{p}_m, \vec{H}_0$ to the right-hand triple (and vice versa). This fact explains mutual orientation of vectors in Figs. 5 b,c.

To realize quantum gates based on coupled ME particles a set of two normally magnetized ferrite disks should be placed in a maximum of RF magnetic field in a cavity. To control the states of every particle different mechanisms can be used. One of examples of a quantum gate is shown in Fig. 6. Every particle is also placed in a region of a maximum RF electric field in a slotline-waveguide resonator (SWR). Oscillations in SWRs are excited due to electrostatic coupling between a SWR and a surface electrode of a particle. The magnetic dipole-dipole interaction between two particles takes place due to MS-potential fields outside ferrite disks.

Quantum gates can be realized with use of collective normal modes of $N$ linearly placed ME particles. The coupling is provided by the magnetic dipole-dipole interaction. Due to the presence of surface electrodes we have the possibility of individually addressing single ME particle (qubit). In this system the measurement (readout of the quantum register) can be carried out with a high efficiency by interaction individual ME particle with different stripline waveguides in standing wave (maximum RF electric field) configurations. A planar construction of such a quantum gate is shown in Fig. 7. In Fig. 7 one can see capacitance connections between surface electrodes and controlling microstrip lines. The model is based upon operation at the node of a standing-wave microstrip-line field.

In a system shown in Fig. 7, excitation of collective ME normal modes is realized with use of two terminal ME particles connected with SWRs. In this case one does not need to place a system in a maximum of RF magnetic field in a cavity. So the construction has a clear technological adventure.

In their seminal work Cirac and Zoller [24] proposed to pursue a mixed strategy in which the quantum information is stored in metastable atomic states and the light coupling to additional auxiliary states is used to perform the quantum gates. To a certain extent the above scheme of a solid-state system resembles the Cirac and Zoller's scheme of quantum computation with cold ions. In our case, for a given bias magnetic field every particle is "cooled" to its lowest quantum state. To modify the state of a single particle (a single qubit) we use local loop currents. Under the particles we place in-plane small wire loops. When a certain-amplitude and certain-direction pulse current in a loop is thrown, a local bias magnetic field of the particle is properly changed to get a two-level transition. The transition between two levels takes place if the operation time is much smaller than the decoherence time. To measure the final qubit states we use controlling microstrip lines.

**(C) Method of ME-particle symmetry transformation**

In the above consideration we discussed that magnetic coupling takes place only for ME particles with the same type of symmetry. However, the fact that we may have different-symmetry (left-hand or right-hand) ME particles gives us an additional degree of freedom. As the third technology, let us consider the possibility to realize a transformation from the left-hand triple of vectors $\vec{p}_e, \vec{p}_m, \vec{H}_0$ to the right-hand triple (and vice versa). To get controllable magnetic coupling between neighboring ME particles, MSW-waveguide sections [13] can be used (see Fig. 8). The particle placed in standing wave configurations of stripline and MSW waveguides interacts, respectively, with external RF electric and magnetic fields (Fig 8 a). Since the stripline and MSW waveguides are completely uncoupled, one can govern an arbitrary phase shift between RF electric and magnetic fields acting on a ME particle. By in-phase and out-of-phase shifts, one can easily realize the left-hand or right-hand configurations of ME fields (Fig. 8 b).

## 6. CONCLUSION

We have demonstrated that based on confinement phenomena for MS oscillations in normally magnetized ferrite disk with discrete-energy-level atomic properties and quantized ME oscillations (resulting from the quantized MS-oscillation properties), the basic elements of quantum computers – the qubits – can be realized. For creation of quantum gates based on MS/ME particles different technologies can be used. One of the attractive features of proposed tecnologies is the possibility of individually addressing single particle (qubit). In such systems the measurement (readout of the quantum register) can be carried out with a high efficiency.

The fact that ME particles can be coupled with stripline or/and MSW waveguides shows a perspective of realization of quantum communications between distant "nodes" over channels. This can be considered as a microwave analog of a quantum-optical model with transmission of a quantum state between two distant atoms via photons [25].

## ACKNOWLEDGEMENT

This paper appears due to scientific collaboration during one of the authors' (E.K.) recent stay in National Center for Theoretical Sciences (NCTS), Hsinchu, Taiwan. E.K. is grateful to NCTS for invitation and hospitality.

FIGURE CAPTIONS

Fig1 ME particle and symmetry properties of ME fields
(a) A view of ferrite ME particle with a linear electrode and capacitive dumb-bells; the left-hand or the right-hand triples of vectors $\vec{p}_e, \vec{p}_m$ and $\vec{H}_0$.
(b) PT transformation of a triple of vectors $\vec{p}_e, \vec{p}_m$ and $\vec{H}_0$.
(c) CPT transformation of a triple of vectors $\vec{p}_e, \vec{p}_m$ and $\vec{H}_0$.

Fig. 2 Mutual matching between the frequency and magnetic-field spectrums.

Fig. 3 ME particle placed in a center of a high-$Q$ rectangular cavity.

Fig. 4 A scheme illustrating the transfer of ME-particle-state coherence to the cavity mode

Fig. 5 Two laterally coupled MS/ME particles
(a) A model of open ferrite disks neglecting the magnetostatic fields in corner regions.
(b) A triple of vectors $\vec{p}_e, \vec{p}_m$ and $\vec{H}_0$ for an even-parity state.
(c) A triple of vectors $\vec{p}_e, \vec{p}_m$ and $\vec{H}_0$ for an odd-parity state.

Fig.6 Two magnetically coupled ME particles with slot-waveguide resonators.

Fig. 7 A quantum gate based on $N$ coupled ME particles operating at the nodes of a standing-wave microstrip-line field.

Fig. 8 Method of ME-particle symmetry transformation.
(a) Stripline and MSW waveguides interacting with a ME particle.
(b) In-phase and out-of-phase shifts between RF electric and magnetic fields.

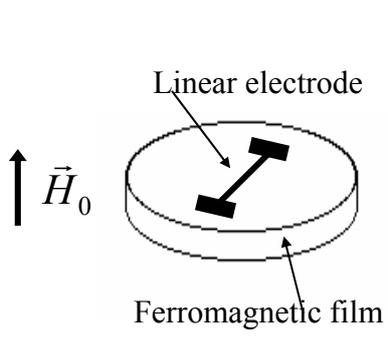
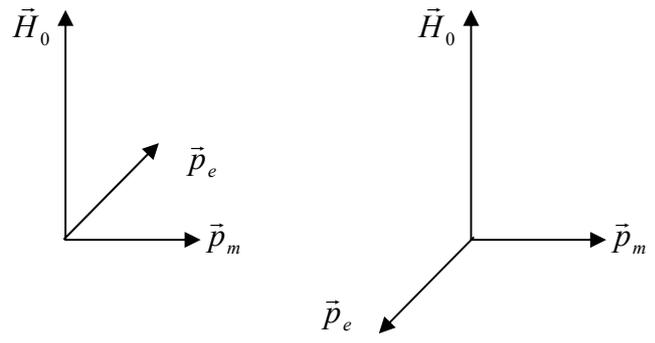

(a)

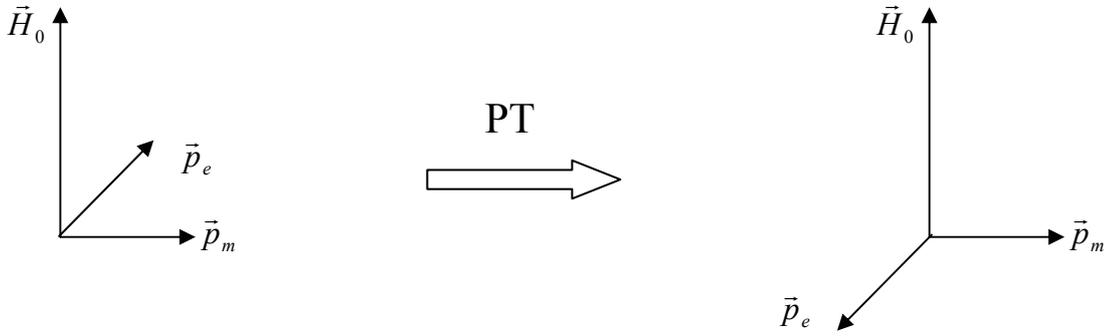

(b)

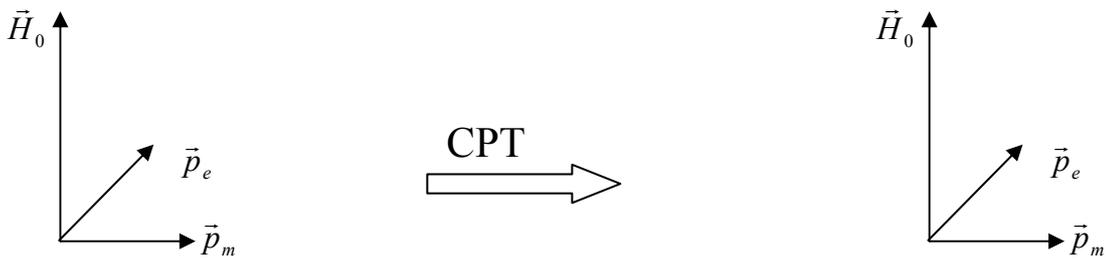

(c)

Fig. 1

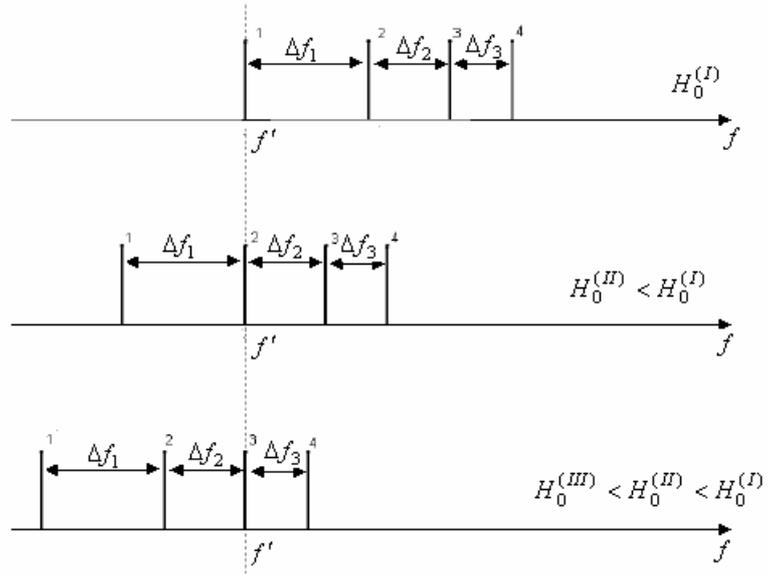

Fig. 2

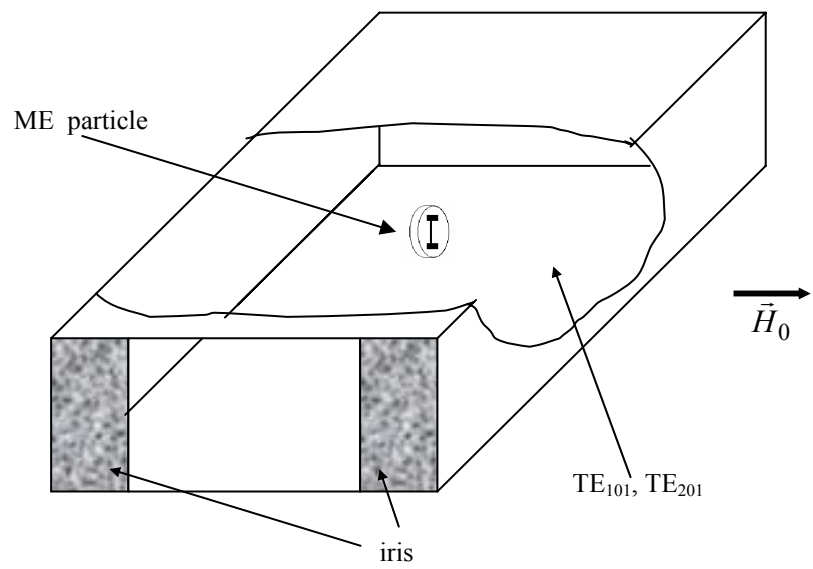

Fig. 3

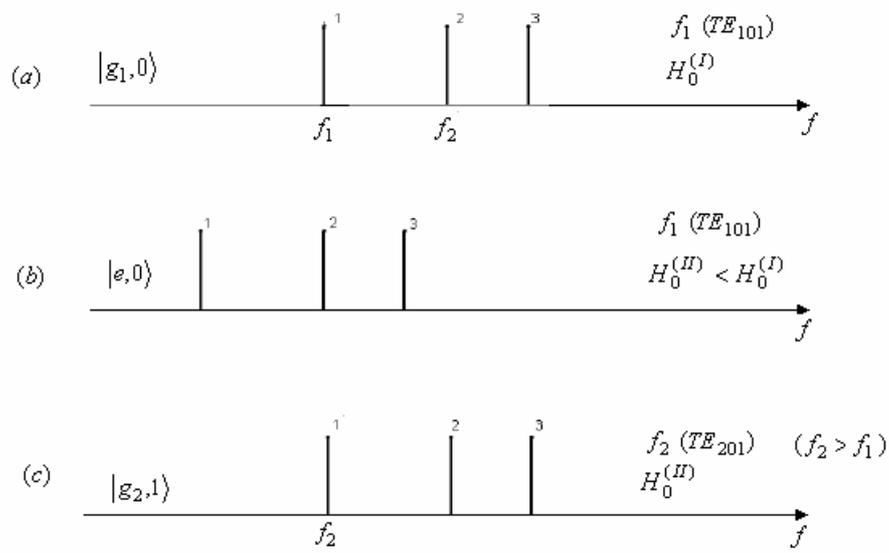

Fig. 4

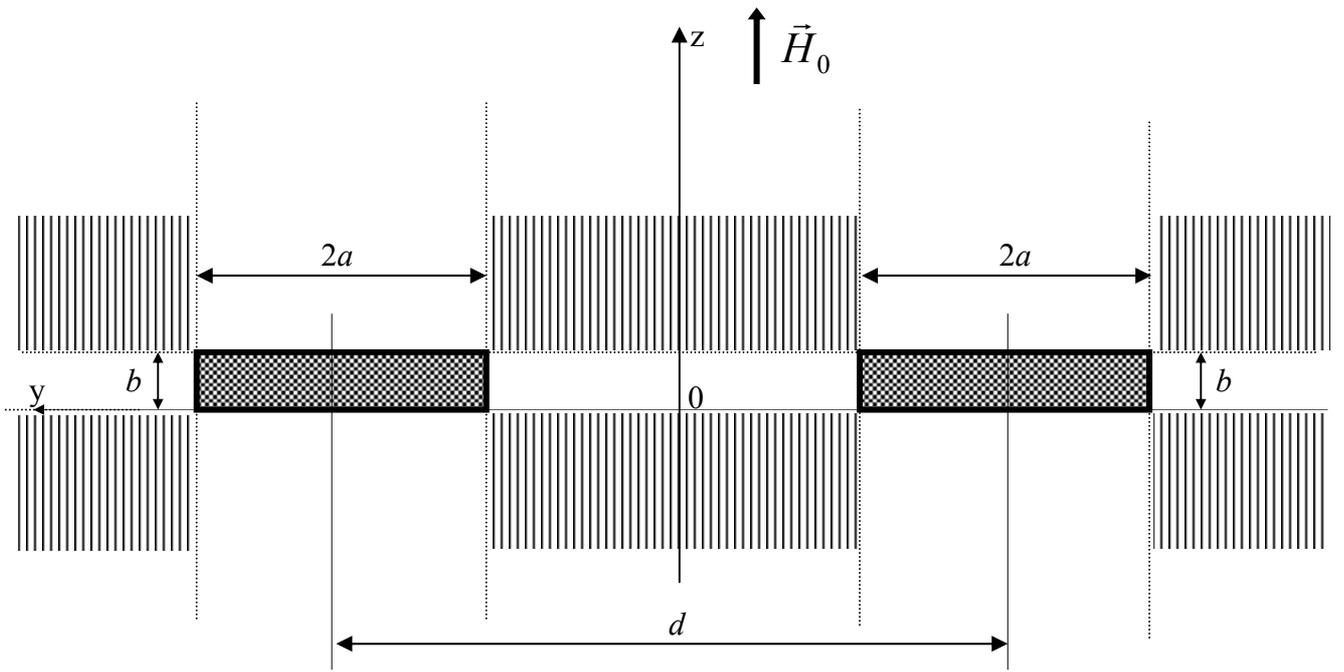

(a)

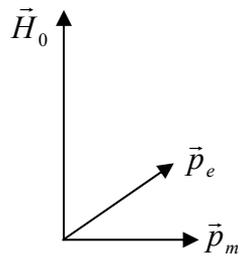 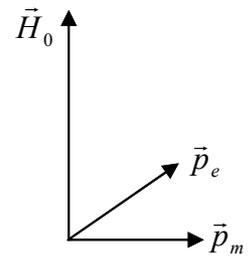

(b)

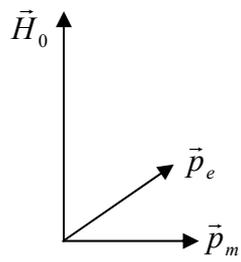 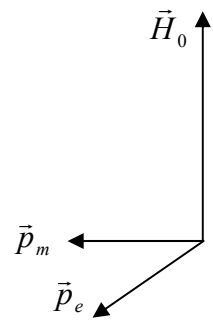

(c)

Fig. 5

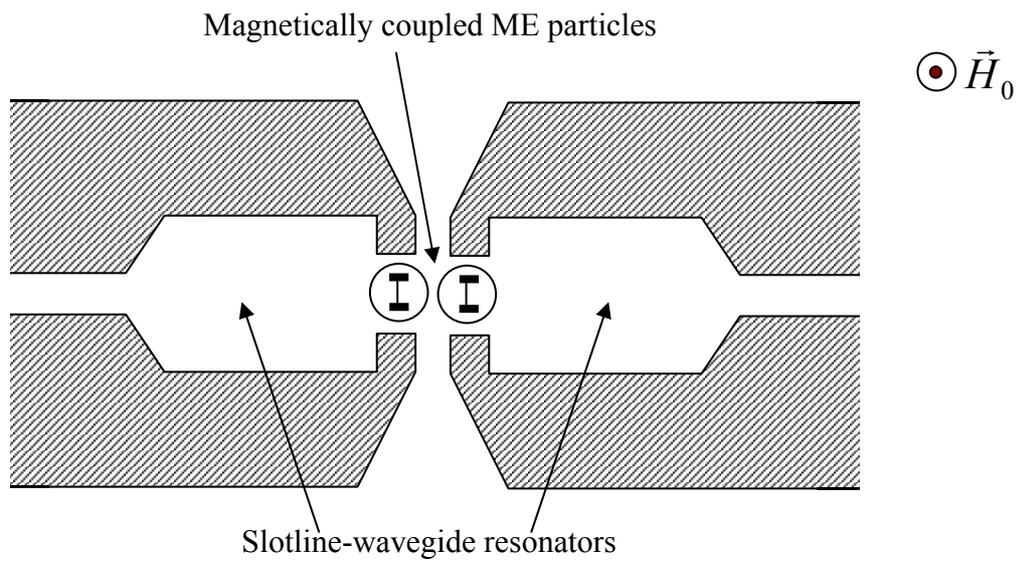

Fig. 6

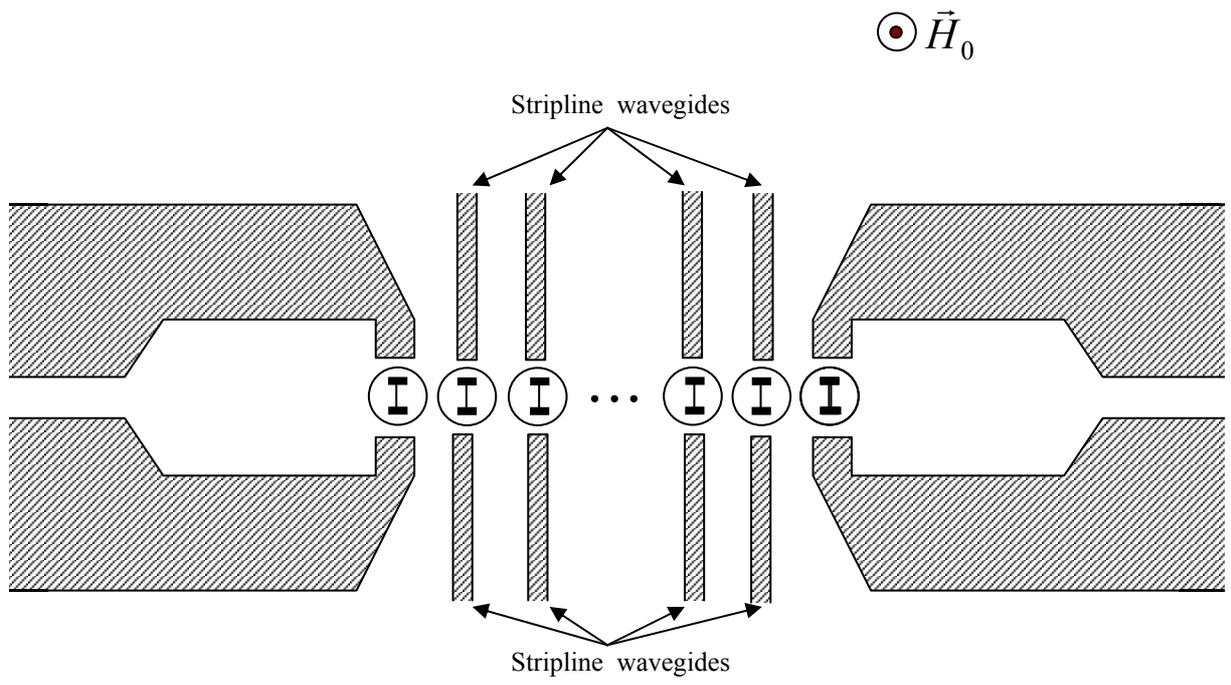

Fig. 7

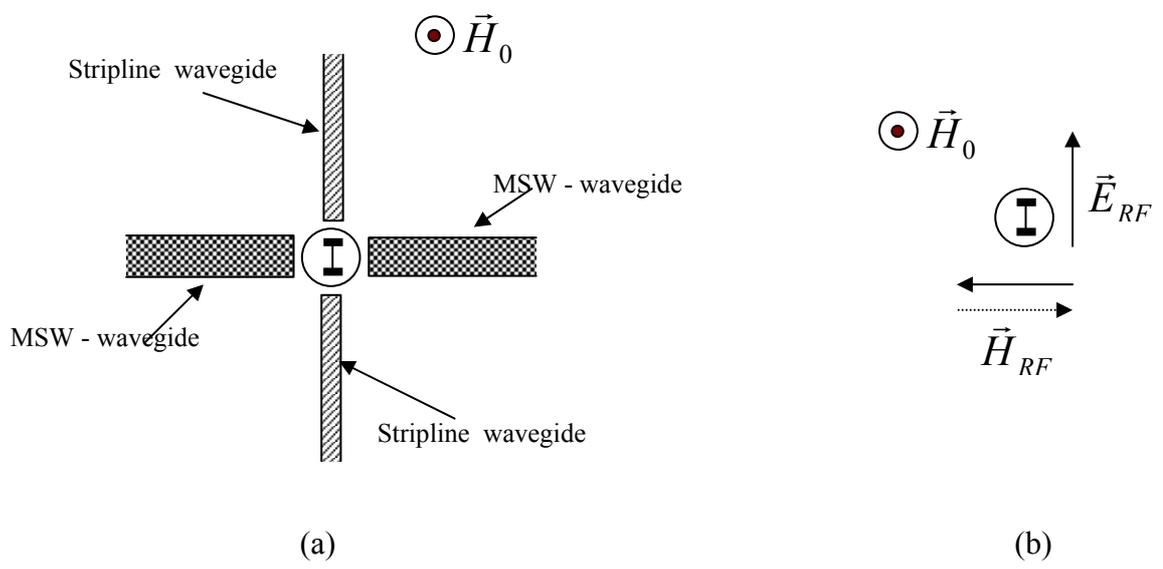

Fig. 8